\documentclass[letterpaper, 10 pt, conference]{ieeeconf}

\IEEEoverridecommandlockouts

\usepackage{amssymb,latexsym,amsfonts,amsmath}

\usepackage{tikz}
\usepackage[normalem]{ulem}
\let\labelindent\relax

\usetikzlibrary{calc,shapes,arrows,intersections,automata,shapes,chains}
\usetikzlibrary{shapes.multipart}
\usepackage{amssymb,latexsym,amsfonts,amsmath}
\usepackage{graphicx}
\usepackage{enumitem}
\usepackage{eso-pic}
\usepackage{mathtools}
\usepackage{epstopdf}
\usepackage{lineno}
\allowdisplaybreaks[4]
\usetikzlibrary{positioning,shapes}
\usetikzlibrary{arrows}

\newcommand{\tup}{\textup}

\newtheorem{theorem}{Theorem}
\newtheorem{definition}[theorem]{Definition}
\newtheorem{corollary}[theorem]{Corollary}

\newtheorem{assumption}[theorem]{Assumption}
\newtheorem{proposition}[theorem]{Proposition}
\newtheorem{remark}[theorem]{Remark}

\newcommand{\I}{\mathcal{I}_d}
\newcommand{\TF}{\mathcal{F}}

\newcommand{\op}{{\mathcal H}}
\def\field#1{\mathbb #1}%
\def\R{\field{R}}%
\def\N{\field{N}}%

\newcommand{\SF}{\mathcal{S}}
\newcommand{\Let}{:=}
\newcommand{\KK}{\mathcal{K}_{\infty}}

\usepackage{color}
\usepackage{xspace}

\renewcommand{\emptyset}{{\varnothing}}
\title{\LARGE \bf Compositional Verification of Initial-State Opacity for Switched Systems}

\author{Siyuan Liu, Abdalla Swikir, and Majid Zamani
	\thanks{This work was supported in part by the H2020 ERC Starting Grant AutoCPS (grant agreement No. 804639), the German Research Foundation (DFG) through the grants ZA 873/1-1, the NSF under Grant ECCS-2015403, China Scholarship Council, and the TUM International Graduate School of Science and Engineering (IGSSE).}
	\thanks{S. Liu and A. Swikir are with the Department of Electrical and Computer Engineering, Technical University of Munich, Germany; {\tt\small \{sy.liu, abdalla.swikir\}@tum.de}. M. Zamani is with the Computer Science Department, University of Colorado Boulder, CO 80309, USA. M. Zamani is also with the Computer Science Department, LMU Munich, Germany; {\tt\small majid.zamani@colorado.edu}. 
}}

\begin{document}

	\maketitle
	\pagestyle{empty}
	\thispagestyle{empty}

	\begin{abstract}                         
	 In this work, we propose a compositional framework for the verification of approximate initial-state opacity  for networks of discrete-time switched systems. The proposed approach is based on a notion of approximate initial-state opacity-preserving simulation functions (InitSOPSFs), which characterize how close concrete networks and their finite abstractions are in terms of the satisfaction of approximate initial-state opacity.
	  We show that such InitSOPSFs can be obtained compositionally by assuming some small-gain type conditions and composing so-called local InitSOPSFs constructed for each subsystem separately. Additionally, for switched systems satisfying certain stability property, we provide an approach to construct their finite abstractions together with the corresponding local InitSOPSFs. Finally, the effectiveness of our results is illustrated through an example.
	\end{abstract}
	
	\section{Introduction}
In recent decades, CPSs have become ubiquitous in critical infrastructures and industrial control systems, including power plants, medical devices and smart communities \cite{cardenas2009challenges}. While the increased connectivity between cyber and physical components brings in the benefit of improving systems functionalities, it also exposes CPSs to more vulnerabilities and security challenges. More recently, the world has witnessed numerous cyber-attacks which have led to great losses in people's livelihoods \cite{ashibani2017cyber}. 
	Therefore, ensuring the security of CPSs has become significantly more important.  
	
	In this work, we focus on an information-flow security property, called \emph{opacity}, which characterizes the ability that a system forbids leaking its secret information to a malicious intruder outside the system. Opacity was firstly introduced in \cite{mazare2004using} to analyze cryptographic protocols. Later, opacity was widely studied in the domain of Discrete Event Systems (DESs), see \cite{lafortune2018history} and the references therein. In this context, existing works on the analysis of various notions of opacity mostly apply to systems modeled by finite state automata, which are more suitable for the cyber-layers of CPSs. However, for the physical components, system dynamics are in general hybrid with uncountable number of states. 
	
	There have been some recent attempts to extend the notions of opacity to continuous-space dynamical systems \cite{ramasubramanian2019notions,zhang2019opacity, yin2019approximate, liu2020stochastic}. In \cite{ramasubramanian2019notions}, a framework for opacity was introduced for the class of discrete-time linear systems, where the notion of opacity was formulated as an output reachability property rather than an information-flow one. 
	The results in \cite{zhang2019opacity} presented a formulation of opacity-preserving (bi)simulation relations between transition systems, which allows one to verify opacity of an infinite-state transition system by leveraging its associated finite quotient one. However, the notion of opacity proposed in this work assumes that the outputs of systems are symbols and are exactly distinguishable from each other, thus, is only suitable for systems with purely logical output sets. In a more recent paper \cite{yin2019approximate}, a new notion of \emph{approximate opacity} was proposed to accommodate imperfect measurement precision of physical systems. Based on this, the authors proposed a notion of so-called approximate opacity-preserving simulation relation to capture the closeness between continuous-space systems and their finite abstractions (a.k.a symbolic models) in terms of preservation of approximate opacity.
	
	The recent results in \cite{liu2020stochastic} investigated opacity for discrete-time stochastic control systems using a notion of so-called initial-state opacity-preserving stochastic simulation functions between stochastic control systems and their finite abstractions (a.k.a. finite Markov decision processes).   
	Though promising, the computational complexity of the construction of finite abstractions grows exponentially with respect to the dimension of the state set, and, hence, the existing approaches \cite{zhang2019opacity,yin2019approximate,liu2020stochastic} will become computationally intractable when dealing with large-scale systems.
	
	Motivated by those abstraction-based techniques in \cite{zhang2019opacity,yin2019approximate,liu2020stochastic} and their limitations, this work proposes an approach to analyze approximate initial-state opacity for networks of switched systems by constructing their opacity-preserving finite abstractions compositionally. There have been some recent results proposing compositional techniques for constructing finite abstractions for networks of systems (see the results in \cite{meyer,7403879,SWIKIR2019,swikir2019compositional} for more details). However, the aforementioned compositional schemes are proposed for the sake of controller synthesis for temporal logic properties, and none of them are applicable to deal with security properties. 

	In this paper, we provide a compositional approach to analyze approximate initial-state opacity of a network of switched systems using their finite abstractions. We first define a notion of so-called local approximate initial-state opacity-preserving simulation functions (InitSOPSFs) to relate each switched system and its finite abstraction. Then, by leveraging some small-gain type conditions, we construct an InitSOPSF as a relation between the network of switched systems and that of their finite abstractions using local InitSOPSFs. This InitSOPSF characterizes the closeness between the two networks in terms of the preservation of approximate initial-state opacity. 
	Moreover, under some assumptions ensuring incremental input-to-state stability of discrete-time switched systems,
	we provide an approach to construct their finite abstractions together with their local InitSOPSFs. Finally, an illustrative example is presented to show how one can leverage our compositionality results for the verification of opacity for a network of switched systems.
	
	Due to lack of space, we provide the proofs of all statements in an arXiv version of the paper \cite{liu2020verification}.
	
	\section{Notation and Preliminaries}\label{1:II}
\emph{Notation}:We denote by $\R$ and $\N$ the set of real numbers and non-negative integers,  respectively.
	These symbols are annotated with subscripts to restrict them in
	the obvious way, e.g. $\R_{>0}$ denotes the positive real numbers. We denote the closed, open, and half-open intervals in $\R$ by $[a,b]$,
	$(a,b)$, $[a,b)$, and $(a,b]$, respectively. For $a,\!b\!\in\!\N$ and $a\!\le\! b$, we
	use $[a;b]$, $(a;b)$, $[a;b)$, and $(a;b]$ to
	denote the corresponding intervals in $\N$.  Given any $a\!\in\!\R$, $\vert a\vert$ denotes the absolute value of $a$.
	Given $N\!\!\in\!\!\N_{\ge1}$ vectors $\nu_i\!\!\in\!\R^{n_i}\!$, $n_i\!\in\!\!\N_{\ge\!1}$, and $i\!\in\!\![1;\!N]$, we
	use $\nu\!=\![\nu_1;\!\ldots\!;\nu_N]$ to denote the vector in $\R^n$ with
	$n\!=\!\sum_i\! n_i$ consisting of the concatenation of vectors~$\nu_i$. Moreover, $\Vert \nu\Vert$ denotes the infinity norm of $\nu$.
	The individual elements in a matrix $A\!\in\! \R^{m\!\times\! n}$, are denoted by $\{A\}_{i,j}$, where $i\!\in\![1;m]$ and $j\!\in\![1;n]$. We denote the zero matrix in $\R^{n\!\times\! n}$ by $0_n$. 
	We denote by $\text{card}(\cdot)$ the cardinality of a given set and by $\emptyset$ the empty set. 
	For any set \mbox{$S\subseteq\R^n$} of the form of finite union of boxes, e.g., $S\!=\!\bigcup_{j=1}^MS_j$ for some $M\in\N$, where $S_j\!=\!\prod_{i=1}^{n} [c_i^j,d_i^j]\!\subseteq\! \R^{n}$ with $c^j_i<d^j_i$, we define $\emph{span}(S)\!=\!\min_{j=1,\ldots,M}\eta_{S_j}$ and $\eta_{S_j}\!=\!\min\{|d_1^j-c_1^j|,\ldots,|d_{n}^j-c_{n}^j|\}$. 
	Moreover, for a set in the form of $X\!\!=\! \!\prod_{i=1}^N X_i$, where $X_i \!\subseteq \!\R^{n_i}$, $\forall i\in[1;N]$, are of the form of finite union of boxes, and any positive (component-wise) vector $\phi \!=\! [\phi_1;\dots;\phi_N]$ with $\phi_i \!\leq \!\emph{span}(X_i)$, $\forall i\in [1;N]$, we define $[X]_\phi\!\!=\!\! \prod_{i=1}^N [X_i]_{\phi_i}$, where $[X_i]_{\phi_i} \!\!=\!\! [\R^{n_i}]_{\phi_i}\cap{X_i}$ and  $[\R^{n_i}]_{\phi_i}\!=\!\{a\in \R^{n_i}\mid a_{j}\!=\!k_{j}\phi_i,k_{j}\in\mathbb{Z},j\!=\!1,\ldots,n_i\}$.
	Note that if $\phi \!=\! [\eta;\dots;\eta]$, where $0<\eta\leq\emph{span}(S)$, we simply use notation $[S]_{\eta}$ rather than $[S]_{\phi}$.
	We use notations $\mathcal{K}$ and $\mathcal{K}_\infty$
	to denote different classes of comparison functions, as follows:
	$\mathcal{K}\!=\!\{\alpha\!:\!\mathbb{R}_{\geq 0} \!\!\rightarrow\!\! \mathbb{R}_{\geq 0}|$ $\alpha$ is continuous, strictly increasing, and $\alpha(0)\!=\!0\}$; $\mathcal{K}_\infty\!=\!\{\alpha \in \mathcal{K}|$ $ \lim\limits_{r \!\rightarrow\! \infty} \alpha(r)\!=\!\infty\}$.
	For $\alpha,\gamma \!\in\! \mathcal{K}_{\infty}$ we write $\alpha\!\le\!\gamma$ if $\alpha(r)\!\le\!\gamma(r)$, and, with abuse of the notation, $\alpha\!=\!c$ if $\alpha(r)\!=\!cr$ for all $c,r\!\geq\!0$. Finally, we denote by $\I$ the identity function over $\R_{\ge0}$, that is $\I(r)\!=\!r, \forall r\!\in\! \R_{\ge0}$.
	Given sets $X$ and $Y$ with $X\!\subset\! Y$, the complement of $X$ with respect to $Y$ is defined as $Y \!\backslash X \!=\! \{x :\! x \!\in\! Y, x \!\notin\! X\}.$  
	\subsection{Discrete-Time Switched Systems} 
	In this work we study discrete-time switched systems of the following form.
	\begin{definition}\label{dtss}
		A discrete-time switched system (dt-SS) $\Sigma$ is defined by the tuple $\Sigma=(\mathbb X,P,\mathbb W,F,\mathbb Y,h)$,
		where 
		\begin{itemize}[leftmargin=*]
			\item $\mathbb X\subseteq\R^n, \mathbb W\subseteq\R^m$, and $\mathbb Y\subseteq\R^q$ are the state set, internal input set, and output set, respectively;
			\item $P=\{1,\dots,m\}$ is the finite set of modes;
			\item $F=\{f_1,\dots,f_m\}$ is a collection of set-valued maps $f_p: \mathbb X\times \mathbb W\rightrightarrows\mathbb X $ for all $p\in P$;
			\item $h: \mathbb X \rightarrow \mathbb Y $ is the output map.
		\end{itemize} 
		The dt-SS $\Sigma $ is described by difference inclusions of the form
		 \begin{align}\label{eq:2}
			\Sigma:\left\{
			\begin{array}{rl}
			\mathbf{x}(k+1)&\!\!\!\!\in f_{\mathsf{p}(k)}(\mathbf{x}(k),\omega(k)),\\
			\mathbf{y}(k)&\!\!\!\!=h(\mathbf{x}(k)),
			\end{array}
			\right.
			\end{align}where $\mathbf{x}\!:\!\mathbb{N}\!\!\rightarrow\!\! \mathbb X $, $\mathbf{y}\!:\!\mathbb{N}\!\!\rightarrow\!\! \mathbb Y$, $\mathsf{p}\!:\!\mathbb{N}\!\!\rightarrow\!\! P$, and $\omega\!:\!\mathbb{N}\!\!\rightarrow\!\! \mathbb W$ are the state, output, switching, and internal input signal, respectively.
		%
		Let $\varphi_k, k \!\in\! \N_{\ge1}$, denote the time when the $k$-th switching instant occurs. 
		We assume that signal $\mathsf{p}$ satisfies a dwell-time condition \cite{liberzon} (i.e. there exists $k_d \!\in\! \N_{\ge1}$, called the dwell-time, such that for all consecutive switching time instants $\varphi_k,\varphi_{k+1}$, $\varphi_{k+1}\!-\!\varphi_{k}\!\geq \!k_d$).	
		If for all $x\!\in\!  \mathbb X, p\!\in\!  P,  w \!\in \! \mathbb W$, $\text{card}(f_p(x,w))\!\leq\!1$ we say the system $\Sigma$ is deterministic, and non-deterministic otherwise. System $\Sigma$ is called finite if $ \mathbb X, \mathbb W$ are finite sets and infinite otherwise. Furthermore, if for all $x\!\in \! \mathbb X$ 
		there exist $ p\!\in\!  P$ and $ w \!\in\!  W $ such that $\text{card}(f_p(x,w))\!\neq\!0$ we say the system is non-blocking. In this paper,
		we assume that all systems are non-blocking.
		
		Note that in this work, we consider switched systems with some secret states. Hereafter, we slightly modify the formulation in Definition \ref{dtss} to accommodate for initial and secret states, as $\Sigma\!=\!(\mathbb X,\mathbb X_0,\mathbb X_s,P,\mathbb W,F,\mathbb Y,h)$, where $\mathbb X_0,\mathbb X_s \subseteq \mathbb X$ are the sets of initial and secret states, respectively. 
	\end{definition}	
	\subsection{Transition Systems}\label{I}
	
	In this section, we employ the notion of transition systems, introduced in \cite{katoen08}, to provide an alternative description of switched systems that can be later directly related to their finite abstractions.
	\begin{definition}\label{tsm} Given a dt-SS $\Sigma\!=\!(\mathbb X,\mathbb X_0,\mathbb X_s,P,\mathbb W,F,\mathbb Y,$ $h)$, we define the associated transition system $T(\Sigma)\!=\!(X,X_0,$ $X_s,U,W,\TF,Y,\op)$ 
		where:
		\begin{itemize}[leftmargin=*]
			\item  $X\!=\!\mathbb X\times P\times\! \{0,\!\dots\!,k_d-1\}$ is the state set; 
		$X_0\!=\!\mathbb X_0\!\times\! P\!\times\! \{0\}$ is the initial state set; 
		$X_s\!=\!\mathbb X_s\!\times\! P\!\times\! \{0,\!\dots\!,k_d-1\}$  is the secret state set; 
			\item $U\!=\!P$ is the external input set;
		$W\!=\!\mathbb{W}$ is the internal input set;
			$Y\!=\!\mathbb{Y}$ is the output set;
		$\mathcal{H}\!:\!X\!\rightarrow\! Y$ is the output map defined as $\mathcal{H}(x,p,l)\!=\!h(x)$;
			\item $\TF$ is the transition function given by $(x^+\!,p^+\!,l^+\!)\!\in\! \TF((x,p,l),u,w)$ if and only if  $x^+\!\in\! f_p(x,w),u\!=\!p$ and the following scenarios hold:
			\begin{itemize} 
				\item$l<k_d-1$, $p^+=p$ and $l^+=l+1$: switching is not allowed because the time elapsed since
				the latest switch is strictly smaller than the dwell time;
				\item $l=k_d-1$, $p^+=p$ and $l^+=k_d-1$: switching is allowed but no switch occurs;
				\item $l=k_d-1$, $p^+\neq p$ and $l^+=0$: switching
				is allowed and a switch occurs.
			\end{itemize}
		\end{itemize}
	\end{definition}
The following proposition is borrowed from \cite{swikir2019compositional} showing that the output runs of a dt-SS $\Sigma$ and its associated transition system $T(\Sigma)$ are equivalent so that one can use $\Sigma$ and $T(\Sigma)$ interchangeably.
	\begin{proposition}\label{traj}
	Consider a transition system $T(\Sigma)$ in Definition \ref{tsm} associated to $\Sigma$ as defined in \eqref{eq:2}. Any output trajectory of $\Sigma$ can be uniquely mapped to an output trajectory of $T(\Sigma)$ and vice versa.
	\end{proposition}	

Next, let us provide a formal definition of networks of dt-SS (or equivalently, networks of transition systems). 
	\subsection{Networks of Systems}
	Consider $N\!\!\in\!\!\N_{\ge1}$ dt-SS $\Sigma_i\!\!=\!\!(\mathbb X_i,\! \mathbb X_{0_i},\!\mathbb X_{s_i},\! P_i,\!\mathbb W_i,\!F_i,\!\mathbb Y_i,$ $\!h_i)$, $i\!\in\![1;N]$, with partitioned internal inputs and outputs as 
	 \begin{align}\label{eq:int1}  
		w_i &= [w_{i1};\ldots;w_{i(i-1)};w_{i(i+\!1)};\ldots;w_{iN}],\\\label{eq:int2}   
		h_{i}(x_i) &= [h_{i1}(x_i);\ldots; h_{iN}(x_i)],   
		\end{align}with $\mathbb{W}_i\!\!=\!\!\prod_{j=1, j\neq i}^{N} \!\mathbb{W}_{ij}$, $\mathbb Y_i\!\!=\!\!\prod_{j=1}^N \! \mathbb Y_{ij}$, $w_{ij} \!\in\! \mathbb{W}_{ij}$, $y_{ij} \!=\!h_{ij}(x_i)\!\in \!\mathbb Y_{ij}$.
	The outputs $y_{ii}$ are considered as external ones, whereas $y_{ij}$,  with $i\!\neq\! j$, are interpreted as internal ones which are used to construct interconnections between systems. In particular, we assume that $w_{ij}$ equals to $y_{ji}$ if there is connection from system $\Sigma_{j}$ to
	$\Sigma_i$, otherwise we set $h_{ji}\!\equiv \!0$. In the sequel, we denote by $\mathcal{N}_i \!=\! \{j \!\in\![1;N], j\!\neq\! i|h_{ji}\!\neq\! 0\}$ the collection of neighboring systems $\Sigma_j,j\!\in\!\mathcal{N}_i$, that provide internal inputs to system $\Sigma_i$.
	
	Now, we are ready to provide a formal definition of the network consisting of $N\!\in\!\N_{\ge1}$ dt-SS.
	\begin{definition}\label{netsw}
		Consider $N\!\in\!\N_{\ge1}$ dt-SS $\Sigma_i\!=\!(\mathbb X_i,\! \mathbb X_{0_i},\!\mathbb X_{s_i},$ $\! P_i,\!\mathbb W_i,\!F_i,\!\mathbb Y_i,h_i)$, $i\!\in\![1;N]$, with the input-output structure given by $\eqref{eq:int1}$ and $\eqref{eq:int2}$. The network, representing the interconnection of $N$ dt-SS $\Sigma_i$, is a tuple $\Sigma\!=\!(\mathbb X,\!\mathbb X_0,\!\mathbb X_s,\!P,\!F,\!\mathbb Y,\!h)$, denoted by $\mathcal{I}_{\mathcal{M}}\!(\Sigma_1,\!\ldots\!,\Sigma_N)$, where $\mathbb X \!=\!\prod_{i=1}^N \mathbb X_i$, $\mathbb X_0 \!=\!\prod_{i=1}^N \mathbb X_{0_i}$, $\mathbb X_s\! =\!\prod_{i=1}^N \mathbb X_{s_i}$,
		$P\!=\!\prod_{i=1}^N \! P_i$, ${F}\!=\!\prod_{i=1}^N\!{F}_i$,  $ \mathbb Y\!=\!\prod_{i=1}^N \! \mathbb Y_{ii}$, 
	$h\!=\!\prod_{i=1}^N \!h_{ii}$, and
	 $\mathcal{M} \!\in\! \mathbb{R}^{N \!\times\! N}$ is a matrix with elements $\{\mathcal{M}\}_{ii} \!=\! 0,\{\mathcal{M}\}_{ij} \!=\! \phi_{ij}, \forall i,\!j \!\in\! [1;N], i\!\neq\! j $, $0\!\leq\!\phi_{ij}\!\leq\! \emph{span}(\mathbb{Y}_{ji})$,
	subject to the constraint:
				\begin{align}\label{const}
	   \Vert y_{ji}\!-\!w_{ij} \Vert \!\leq\! \phi_{ij}, [\mathbb{Y}_{ji}]_{\phi_{ij}} \!\subseteq\! \mathbb{W}_{ij}, \forall i\!\in\! [1;N], j\!\in\!\mathcal{N}_i. 
		\end{align} 
	\end{definition}
	\begin{remark}
	In this paper, when we are talking about the network of concrete switched systems, $y_{ji}$ is always equal to $w_{ij}$, which naturally implies $\phi_{ij}=0$ and $\mathcal{M}=0_N$. However, for the network of finite abstractions, due to possibly different granularities of 
	the internal input and output sets, the designed parameters ${\phi_{ij}}$ are not necessarily zero.	 
	Note that whenever ${\phi_{ij}}\!\neq\! 0$, the sets $\mathbb{{Y}}_{ji}$, $\forall i,\!j\!\in\![1;N],~i\!\neq\! j$, are assumed to be finite unions of boxes.
\end{remark}

Similarly, given transition systems $T(\Sigma_i)$, one can also define a network of transition systems $\mathcal{I}_{\mathcal{M}}(T(\Sigma_1),\!\ldots\!,T(\Sigma_N))$.
	\section{Opacity-preserving Simulation Functions}\label{I}
	In this section, we start by defining approximate initial-state opacity property \cite{yin2019approximate} for networks of transition systems. This property is, in general, hard to check for a concrete network as its state set is infinite and so far there is no systematic way in the literature to verify opacity of such systems. On the other hand, existing tool DESUMA\footnote{Available at URL http://www.eecs.umich.edu/umdes/toolboxes.html.} and algorithms \cite{yin2017new},\cite{saboori2013verification},\cite[Sec. IV]{zhang2019opacity} in DESs literature can be leveraged to
	check opacity for systems with finite state sets. Therefore, it would be feasible to check opacity for finite networks (i.e, networks consisting of finite abstractions) and then carry back the reasoning to concrete ones, as long as there is a formal relation between those networks. To this purpose, we introduce a new notion of approximate initial-state opacity preserving simulation functions (InitSOPSF) which formally relate two networks of transition systems and their approximate initial-state opacity properties. 
	
	Before defining the notion of approximate initial-state opacity for networks of transition systems, we introduce some notations as follows. 
	Consider network $T(\Sigma)$. 
	We use $z^k$ to denote the state of $T(\Sigma)$ reached at time $k \!\in\! \mathbb{N}$ from the initial state $z^0$ under the input sequence ${\bar u}$ with length $k$, and denote by $\{z^0\!,\! z^1\!, \!\dots\!,\! z^k\!\}$ a finite state run of $T(\Sigma)$ with length $k\!\in\! \mathbb{N}$.
	
	\begin{definition}
		Consider network $T(\Sigma)\!=\!\!(X,X_0,X_s,U,\TF,$ $Y,\op)$  
		and a constant $\delta \!\geq\! 0$. Network $T(\Sigma)$ is said to be
			$\delta$-approximate initial-state opaque if for any $z^0 \!\in\! X_0 \!\cap\! X_s$ and finite state run $\{z^0\!, z^1\!,\! \dots\!, z^k\!\}$, there exist $\bar{z}^0\! \in\! X_0 \!\setminus \!X_s$ and a finite state run $\{\bar{z}^{0}\!, \bar{z}^{1}\!, \!\dots\!, \bar{z}^{k}\!\}$ such that 	
 \begin{align*}
			\max_{t \in [0;k]} \Vert \mathcal{H}(z^t)-\mathcal{H}(\bar{z}^t) \Vert \leq \delta.
	\end{align*}
	\end{definition}
	Now, we can introduce a notion of approximate InitSOPSF to quantitatively relates two networks of transition systems in terms of preservation of approximate opacity as defined above.
	
	\begin{definition}\label{sfg}
		Consider $T(\Sigma)\!=\!\!(X,\!X_0,\!X_s,\!U,\!\TF,\!Y,\!\op)$ and $T(\hat{\Sigma})\!=\!(\hat{X},\!\hat X_0,\!\hat X_s,\!\hat{U},\!\hat{\TF},\!\hat{Y},\!\hat{\op})$ with $\hat Y\!\subseteq\! Y$. For $\varepsilon \!\in\! \mathbb{R}_{\geq0}$, a function ${\SF}\!:\! X\!\times\! \hat X \!\to\! \mathbb{R}_{\geq0} $ is called an $\varepsilon$-approximate initial-state opacity-preserving simulation function ($\varepsilon$-InitSOPSF) from ${T}(\Sigma)$ to $T(\hat{\Sigma})$ if there exists $\alpha \!\in\! \mathcal{K_{\infty}}$ such that 
	 \begin{itemize}[leftmargin=*]
				\item[1](a) $\forall z^0 \in {X}_0 \cap {X}_s$, $\exists \hat z^0 \in \hat {X}_0 \cap \hat {X}_s$, s.t. ${\SF}(z^0,\hat z^0) \leq \varepsilon $;\\
				(b) $\forall \hat z^0 \in \hat {X}_0 \setminus \hat {X}_s$, $\exists z^0 \in {X}_0 \setminus {X}_s$, s.t. ${\SF}(z^0,\hat z^0) \leq \varepsilon $;
				\item[2]   $\forall z \in X, \forall \hat z \in \hat {X}$, $\alpha(\Vert \mathcal{H}(z) - \mathcal{\hat H}(\hat z) \Vert) \leq {\SF}(z,\hat z)$;
				\item[3] $\forall z \!\in\! X, \forall \hat z \!\in\! \hat {X}$ s.t. $\mathcal{S}(z,\hat z) \!\leq \!\varepsilon$, one has: \\
				(a) $\forall u\in  U$, $\!\forall z^+\!\!\in\!\TF(z,u)$, $\exists \hat u\in \hat U$, $\exists \hat z^+\! \!\in\!\hat{\TF}(\hat{z},\hat{u})$, s.t. ${\SF}(z^+\!\!,\hat z^+\!) \!\leq\! \varepsilon$;\\
				(b) $\forall \hat u\in \hat U$, $\!\forall \hat z^+\!\! \in\! \hat{\TF}(\hat{z},\hat{u})$, $\exists u\in U$, $\exists z^+ \!\!\in\! \TF(z,u)$,  s.t. ${\SF}(z^+\!\!,\hat z^+\!)\! \leq\! \varepsilon$.
		\end{itemize}
	\end{definition}

	Although Definition \ref{sfg} is general in the sense that networks $T(\Sigma)$ and $T(\hat{\Sigma})$ can be either infinite or finite, practically, network $T(\hat{\Sigma})$ potentially consists of $N\!\!\in\!\!\N_{\ge1}$ finite abstractions. Hence, checking approximate initial-state opacity for this network is decidable  in comparison to network $T(\Sigma)$. 
	
	Before showing the next result, let us recall the definition of approximate initial-state opacity-preserving simulation relation which was originally proposed in \cite{yin2019approximate}.
	\begin{definition}\label{def:InitSOP}
		Consider networks $T(\Sigma)\!=\!(\!X,\!X_0,\!X_s,\!U,$ $\!\TF,\!Y,\!\op)$ and $T(\hat{\Sigma})\!=\!(\!\hat{X},\!\hat X_0,\!\hat X_s,\!\hat{U},\!\hat{\TF},\!\hat{Y},\!\hat{\op})$ where $\hat Y\!\subseteq \!Y$. For $\hat{\varepsilon} \!\in\! \mathbb R_{\ge 0}$, a relation  $R\subseteq X\times\! \hat{X}$  is called an  $\hat{\varepsilon}$-approximate initial-state opacity-preserving simulation relation ($\hat{\varepsilon}$-InitSOP simulation relation) from ${T}(\Sigma)$ to $T(\hat{\Sigma})$ if 
	\begin{itemize}[leftmargin=*]
		\item[1] (${a}$) $\forall z^0 \in {X}_0 \cap {X}_s$, $\exists \hat z^0 \in \hat {X}_0 \cap \hat {X}_s$, s.t.	$(z^0,\hat z^0) \in R$;\\
		(${b}$) $\forall \hat z^0 \in \hat {X}_0 \setminus \hat {X}_s$, $\exists z^0 \in {X}_0 \setminus {X}_s$, s.t. $(z^0,\hat z^0) \in R$;
		\item[2]  $\forall (z, \hat{z}) \in R$, $\Vert \mathcal{H}(z) - \mathcal{\hat H}(\hat z) \Vert\leq \hat{\varepsilon}$;
		\item[3]  For any $(z, \hat{z}) \in R$, we have\\
		(${a}$) $\!\forall u\!\in\!\!  U$, $\!\forall \!z^+ \!\!\in\!\TF(z,\!u)$, $\!\exists \hat u\!\in\! \hat U$, $\!\exists \hat z^+ \!\!\in\!\hat{\TF}(\hat{z},\!\hat{u})$, s.t. $\!(z^+\!\!,\!\hat z^+) \!\in \!R $; \\
		(${b}$) $\forall \hat u\!\in\!\! \hat U$, $\!\forall \!\hat z^+ \!\!\in\! \hat{\TF}(\hat{z},\!\hat{u})$, $\!\exists u\!\in\!  U$, $\exists z^+\!\! \in\! \TF(z,u)$, s.t. $\!(z^+\!\!,\!\hat z^+\!) \!\in\! R $.
\end{itemize}
	\end{definition}
	The following corollary borrowed from \cite{yin2019approximate} shows the usefulness of Definition \ref{def:InitSOP} in terms of preservation of approximate opacity across related networks.
	\begin{corollary}\label{thm:InitSOP}
		Consider networks $T(\Sigma)\!=\!(\!X\!,X_0,\!X_s,\!U\!,\TF\!,$ $Y\!,\op)$ and $T(\hat{\Sigma})\!=\!(\!\hat{X}\!,\hat X_0,\!\hat X_s,\!\hat{U}\!,\hat{\TF}\!,\hat{Y}\!,\hat{\op})$ where $\hat Y\subseteq Y$. Let $\hat\varepsilon,\delta\!\in\!\mathbb R_{\ge 0}$.
		If  there exists an $\hat{\varepsilon}$-InitSOP simulation relation from ${T}(\Sigma)$ to $T(\hat{\Sigma})$ as in Definition \ref{def:InitSOP} and $\hat{\varepsilon}\!\leq\! \frac{\delta}{2}$,
		then the following implication holds 
		\begin{align}
		&T(\hat{\Sigma})\tup{ is ($\delta\!-\!2\hat{\varepsilon}$)-approximate initial-state opaque} \nonumber \\
		&\Rightarrow {T}(\Sigma) \tup{ is $\delta$-approximate initial-state opaque}.\nonumber
		\end{align}
	\end{corollary}

The next result shows that the existence of an $\varepsilon$-InitSOPSF for networks of transition systems implies the existence of an $\hat{\varepsilon}$-InitSOP simulation relation between them.

\begin{proposition}\label{error}
	Consider networks $T(\Sigma)\!=\!(\!X,\!X_0,\!X_s,\!U,$ $\!\TF,\!Y,\!\op)$  and $T(\hat{\Sigma})\!=\!(\!\hat{X},\!\hat X_0,\!\hat X_s,\!\hat{U},\!\hat{\TF},\!\hat{Y},\!\hat{\op})$ where $\hat Y\!\subseteq\! Y$. Assume ${\SF}$ is an $\varepsilon$-InitSOPSF from ${T}(\Sigma)$ to $T(\hat{\Sigma})$  as in Definition \ref{sfg}. Then, relation $R\!\subseteq\! X\!\times\! \hat{X}$ defined by 
 \begin{align}\label{re}
		R\!=\!\left\{\!(z,\!\hat z)\!\in\! {X}\!\times\! \hat{X}|{\SF}(z,\hat z)\!\leq \!{\varepsilon}\!\right\},  
		\end{align}is an $\hat{\varepsilon}$-InitSOP simulation relation from ${T}(\Sigma)$ to $T(\hat{\Sigma})$ with  
 \begin{align}\label{er}
		\hat{\varepsilon}={\alpha}^{-1}({\varepsilon}).
		\end{align}
\end{proposition}

Given the results of Corollary \ref{thm:InitSOP} and Proposition \ref{error}, one can readily see that if there exists an $\varepsilon$-InitSOPSF from ${T}(\Sigma)$ to $T(\hat{\Sigma})$ as in Definition \ref{sfg} and $T(\hat{\Sigma})$ is ($\delta\!-\!2\hat{\varepsilon}$)-approximate initial-state opaque, then $T(\Sigma)$ is $\delta$-approximate initial-state opaque, where $\hat{\varepsilon}={\alpha}^{-1}({\varepsilon}) \leq  \frac{\delta}{2}, \delta \in \mathbb R_{\ge 0}$.

	\section{Compositionality Result}\label{1:III}
	\label{s:inter}
	We saw in the previous section that $\varepsilon$-InitSOPSFs are very useful for checking approximate initial-state opacity of concrete networks based on checking that of their finite abstractions.  However, constructing such a function for networks consisting of a large number of systems is not feasible in general. Hence, in this section, we introduce a compositional technique based on which one can construct an $\varepsilon$-InitSOPSF from the concrete network to a network of finite abstractions by using so-called local $\varepsilon_i$-InitSOPSFs between subsystems and their abstarctions.

	\subsection{Compositional Construction of $\varepsilon$-InitSOPSF} 
	Suppose that we are given $N$ dt-SS $\Sigma_i$, or equivalently, $T(\Sigma_i)$. Moreover, we assume that systems $T(\Sigma_i)$ and $T(\hat{\Sigma}_i)$ admit a local $\varepsilon_i$-InitSOPSF as defined next.
	\begin{definition}\label{sf}
		Consider $T(\Sigma_i)\!=\!(\!X_i,\!X_{0_i},\!X_{s_i},\!U_i,\!W_i,\!\TF_i,$ $\!Y_i,\!\op_i\!)$ and $T(\hat{\Sigma}_i)\!=\!(\hat{X}_i,\!\hat X_{0_i},\!\hat X_{s_i},\!\hat{U}_i,\!\hat{W}_i,\!\hat{\TF}_i,\!\hat{Y}_i,\!\hat{\op}_i)$, $\forall i\!\!\!\in\!\!\! [1;N]$,  where $\hat W_i\!\subseteq\! W_i$ and $\hat Y_i\!\subseteq\! Y_i$. For $\varepsilon_i \!\in\! \mathbb{R}_{\geq0}$, a function $\mathcal{S}_i \!:\! X_i\!\times\! \hat X_i \!\to \!\mathbb{R}_{\geq0} $ is called a local $\varepsilon_i$-InitSOPSF from $T(\Sigma_i)$ to $T(\hat{\Sigma}_i)$  if there exist $\vartheta_i \!\in\! \mathbb R_{\ge 0}$ and $\alpha_i \!\in\! \mathcal{K_{\infty}}$ such that 
		\begin{itemize}[leftmargin=*]
				\item[1](a) $\forall z^0_i\! \in\! {X}_{0_i} \!\cap\! {X}_{s_i}$, $\exists \hat z^0_i \!\in\! \hat {X}_{0_i} \!\cap\! \hat {X}_{s_i}$, s.t. $\mathcal{S}_i(z^0_i,\hat z^0_i) \!\leq\! \varepsilon_i $;\\
				(b) $\forall \hat z^0_i\!\in \!\hat {X}_{0_i}\! \setminus \!\hat {X}_{s_i}$, $\exists z^0_i \in {X}_{0_i}\!\setminus\! {X}_{s_i}$, s.t. $\mathcal{S}_i(z^0_i,\hat z^0_i) \!\leq\! \varepsilon_i $;
				\item[2]   $\forall z_i \!\in\! X_i, \forall \hat z_i \!\in\! \hat {X}_i$, $\alpha_i(\Vert \mathcal{H}_i(z_i) \!-\! \mathcal{\hat H}_i(\hat z_i) \Vert) \leq \mathcal{S}_i(z_i,\hat z_i)$;
				\item[3] $\forall z_i \!\in\! X_i, \forall \hat z_i \!\in\! \hat {X}_i$ s.t. $\mathcal{S}_i(z_i,\hat z_i) \!\leq\! \varepsilon_i$, $\forall w_i \!\in\! W_i$, $\forall \hat w_i \!\in\! \hat{{W}_i}$ s.t. $\Vert w_i\!-\! \hat w_i \Vert \!\leq\! \vartheta_i$, the following conditions hold: \\
				(a) $\forall u_i\!\in\! U_i$, $\forall z^+_i \!\in\!\TF_i(z_i,u_i,w_i)$, $\exists \hat u_i\!\in\! \hat U_i$, $\exists \hat z^+_i \!\in\!\hat{\TF}_i(\hat{z}_i,\!\hat{u}_i,\!\hat{w}_i)$, s.t. $\mathcal{S}_i(z^+_i\!,\hat z^+_i\!) \!\leq\! \varepsilon_i$;\\
				(b) $\forall \hat u_i\!\in\! \hat U_i$, $\forall \hat z^+_i \!\in\! \hat{\TF}_i(\hat{z}_i,\!\hat{u}_i,\!\hat{w}_i)$, $\exists u_i\!\in\! U_i$, $\exists z^+_i \!\in\! \TF_i(z_i,u_i,w_i)$, s.t. $\mathcal{S}_i(z^+_i\!,\hat z^+_i\!) \!\leq\! \varepsilon_i$.
		\end{itemize}
If there exists a local $\varepsilon_i$-InitSOPSF from $T(\Sigma_i)$ to $T(\hat{\Sigma}_i)$, and $T(\hat{\Sigma}_i)$ is finite ($\hat X_i$ and $\hat W_i$ are finite sets), $T(\hat{\Sigma}_i)$ is called a finite abstraction (or symbolic model) of $T(\Sigma_i)$, which is constructed later in Definition \ref{smm}.
		Note that local $\varepsilon_i$-InitSOPSFs are mainly for constructing an $\varepsilon$-InitSOPSF for the networks and they are not directly used for deducing approximate initial-state opacity-preserving simulation relation.
	\end{definition}
	
	The next theorem provides a compositional approach to construct an $\varepsilon$-InitSOPSF from $T(\Sigma)$ to $T(\hat{\Sigma})$ via local $\varepsilon_i$-InitSOPSFs from $T(\Sigma_i)$ to $T(\hat{\Sigma}_i)$.
	
	\begin{theorem}\label{thm:3}
		Consider network $T(\Sigma)\!=\!\mathcal{I}_{0_N}(T(\Sigma_1),\!\ldots\!,$ $T(\Sigma_{N}))$. Assume that there exists a local $\varepsilon_i$-InitSOPSF $\SF_i$ from $T(\Sigma_i)$ to $T(\hat{\Sigma}_i)$, $\forall i\!\in\! [1;N]$, as in Definition \ref{sf}. Let $\varepsilon \!=\! \max\limits_{i} \varepsilon_i$, 
		and $\mathcal{\hat M} \!\in \!\mathbb{R}^{N\! \times\! N}$ be a matrix with elements $\{\mathcal{\hat M}\}_{ii}\! =\! 0,\{\mathcal{\hat M}\}_{ij}\! =\! \phi_{ij}$, $\forall i,\!j \!\in\! [1;N]$, $i \!\neq\! j$, $0\!\leq\!\phi_{ij}\!\leq\! \emph{span}({\hat Y}_{ji})$.
		If $\forall i \!\in\! [1;N]$, $\forall j \!\in\! \mathcal{N}_i$, 
	 \begin{align} \label{compoquaninit}
			\alpha^{-1}_{j}(\varepsilon_j) + \phi_{ij}\leq \vartheta_i,
			\end{align}
			then, function ${\SF}\!:X\!\times\! \hat{X}\!\rightarrow \!\R_{\ge0}$ defined as
		\begin{align}\label{defVinit}
			{\SF}&(z,\hat{z})\Let\max\limits_{i}\{ \frac{\varepsilon}{\varepsilon_i} \SF_i(z_{i},\hat{z}_{i}) \},
			\end{align}
			is an $\varepsilon$-InitSOPSF from $T(\Sigma)\!=\!\mathcal{I}_{0_N}(T(\Sigma_1),\!\ldots\!,T(\Sigma_{N}))$ to $T(\hat{\Sigma})\!=\!{\mathcal{I}}_\mathcal{\hat M}(T(\hat{\Sigma}_1),\!\ldots\!,T(\hat{\Sigma}_{N}))$.
	\end{theorem}
\begin{remark}
Let $\phi_i = [\phi_{i1};\!\ldots\!;\phi_{iN}]$. Vectors $\phi_i$ serve later as internal input quantization parameters
	for the construction of symbolic models for $T(\Sigma_i)$ (see Definition \ref{smm}). Moreover, the values of $\phi_{i}$ will be designed later in Theorem \ref{smallgain}.
\end{remark}

	\section{Construction of Symbolic Models}\label{1:IV}
	In this section, we consider $\Sigma=(\mathbb X,\mathbb X_0,\mathbb X_s, P, \mathbb W,F,\mathbb Y,h)$ as an infinite, deterministic dt-SS.
	Note that throughout this section, we are mainly talking about switched subsystems rather than the overall network. However, for the sake of better readability, we often omit index $i$ of subsystems throughout the text in this section.
We assume the output map $h$ satisfies the following general Lipschitz assumption: there exists an $\ell\in\KK$ such that: $\Vert h(x)\!-\!h(y)\Vert\!\leq\! \ell(\Vert x\!-\!y\Vert)$ for all $x,y\!\in\! \mathbb X$. Here, we also use $\Sigma_{p}$ to denote a dt-SS $\Sigma$ in \eqref{eq:2} with constant switching signal $\mathsf{p}(k)\!=\!p,~\forall k\!\in\! \N$. 

Here, we establish an $\varepsilon$-InitSOPSF between $T(\Sigma)$ and its symbolic model by assuming that, for all $p \!\in\! P$, $\Sigma_p$ is incrementally input-to-state stable ($\delta$-ISS) \cite{ruffer} as defined next.
	\begin{definition}\label{def:SFD1} 
	System $\Sigma_{p}$ is $\delta$-ISS if there exist functions $ V_p\!:\mathbb X\!\times\! \mathbb X \!\to \!\mathbb{R}_{\geq0} $, $\underline{\alpha}_p, \overline{\alpha}_p, \rho_{p} \!\in\! \mathcal{K}_{\infty}$, and constant $0\!<\!\kappa_p\!<\!1$, such that for all $x,\hat x\!\in\! \mathbb{X}$, and for all $w,\hat w\!\in\! \mathbb{W}$  
		\begin{align}\label{e:SFC11}
		\underline{\alpha}_p (\Vert x&-\hat{x}\Vert ) \leq V_p(x,\hat{x})\leq \overline{\alpha}_p (\Vert x-\hat{x}\Vert ),\\\label{e:SFC22}
		V_p(f_p(x,w),&f_p(\hat x,\hat w))
		\!\leq\! \kappa_p V_p(x,\hat{x})+\rho_{p}(\Vert w- \hat{w}\Vert ).
		\end{align}
	\end{definition}
We say that $V_p$, $\forall p\!\in\! P$, are multiple $\delta$-ISS Lyapunov functions for system $\Sigma$ if it satisfies \eqref{e:SFC11} and \eqref{e:SFC22}. Moreover, if $V_{p}\!=\!V_{q}, \forall p,q\!\in\! P$, we omit the index $p$ in \eqref{e:SFC11}-\eqref{e:SFC22}, and say that $V$ is a common $\delta$-ISS Lyapunov function for system $\Sigma$. 
	
	Now, we show how to construct a symbolic model $T(\hat{\Sigma})$ of $T(\Sigma)$ associated with the dt-SS $\Sigma$.
	\begin{definition}\label{smm} Consider a transition system $T(\Sigma)\!=\!(X,\!X_0,$ $\!X_s,\!U,\!W,\!\TF,\!Y,\!\op)$, associated with the switched system $\Sigma\!=\!(\mathbb X,\!\mathbb X_0,\!\mathbb X_s,\!P,\!\mathbb W, \!F,\!\mathbb Y,h)$, where $\mathbb X$, $\mathbb W$ are assumed to be finite unions of boxes. Let $\Sigma_{p}$, $\forall p\!\in\! P$, be $\delta$-ISS as in Definition \ref{def:SFD1}. Then one can construct a symbolic model $T(\hat{\Sigma})\!=\!(\hat{X},\!\hat X_0,$ $\!\hat X_s,\!\hat{U},\!\hat{W},\!\hat{\TF},\!\hat{Y},\!\hat{\op})$ where:
		\begin{itemize}[leftmargin=*]
			\item $\hat{X}\!=\!\hat{\mathbb{X}}\!\times\! P\!\times\! \{0,\!\dots\!,k_d\!-\!1\}$, where $\hat{\mathbb{X}}\!=\![\mathbb{X}]_{\eta}$ and $0 \!<\! \eta \!\leq\! \tup{min} \{span(\mathbb X_s),span(\mathbb{X} \setminus \mathbb X_s)\}$ is the state set quantization parameter;
		 $\hat X_0\!=\!\hat{\mathbb X}_0\!\times\!  P \!\times \!\{0\}$, where $\hat{\mathbb X}_0 \!=\![\mathbb{X}]_{\eta}$; 
		$\hat X_s\!=\!\hat{\mathbb X}_s\!\times \!P\times\!  \{0,\!\cdots\!,k_d\!-\!1\}$, where $\hat{\mathbb X}_s \!=\! [\mathbb X_s]_{\eta}$; 
			\item $\hat{U}\!=\!U\!=\!P$;
			$\hat{Y}\!=\!Y$;
			$\hat{\op}\!:\!\hat{X}\!\rightarrow\! \hat{Y}$, defined as $\hat{\op}(\hat{x},p,l)\!=\!{\op}(\hat{x},p,l)\!=\!h(\hat{x})$;
		$\hat{W}\!=\![\mathbb{W}]_{{\phi}}$, where $\phi$, satisfying $0 \!<\!\Vert \phi \Vert\!\leq\! span(\mathbb W)$, is the internal input set quantization parameter;
			\item $(\hat{x}^+\!,p^+\!,l^+)\!\in\! \hat{\TF}((\hat{x},p,l),\hat{u},\hat{w})$ if and only if $\Vert f_p(\hat{x},\hat{w})\!-\!\hat{x}^+\Vert\!\leq\! \eta$, $\hat{u}=p$ and the following scenarios hold:
			\begin{itemize}
				\item $l<k_d-1$, $p^+=p$ and $l^+=l+1$;
				\item $l=k_d-1$, $p^+=p$ and $l^+=k_d-1$;
				\item $l=k_d-1$, $p^+\neq p$ and $l^+=0$;
			\end{itemize}
		\end{itemize} 
	\end{definition}
In order to construct a local $\varepsilon$-InitSOPSF from $T(\Sigma)$ to $T(\hat{\Sigma})$, we raise the following assumptions on functions $V_p$ appeared in Definition \ref{def:SFD1}.
	\begin{assumption}\label{ass1} 
		There exists $\mu \geq 1$ such that
		\begin{align}\label{mue}
			\forall x,y \in \mathbb{X},~~ \forall p,q \in P,~~ V_p(x,y)\leq \mu  V_{q}(x,y). 
			\end{align}
	\end{assumption}
Assumption \ref{ass1} is an incremental version of a similar assumption in \cite{VU} that is used to prove input-to-state stability of switched systems under constrained switching signals.
	\begin{assumption}\label{ass2} 
		Assume $\exists$ $\gamma_p\in\mathcal{K}_{\infty}$, $\forall p\!\in\! P$, such that
			\begin{align}\label{tinq} 
			\forall x,y,z \in \mathbb{X},~~V_p(x,y)\leq V_p(x,z)+\gamma_p(\Vert y-z\Vert).
			\end{align}
	\end{assumption}
	Assumption \ref{ass2} is non-restrictive as shown in \cite{zamani2014symbolic} provided that one is interested to work on a compact subset of $\mathbb{X}$.   
	
	Now, we establish the relation between $T(\Sigma)$ and $T(\hat{\Sigma})$ via the notion of local $\varepsilon$-InitSOPSF as in Definition \ref{sf}.
	\begin{theorem}\label{thm:2}
		Consider a dt-SS $\Sigma\!=\!(\mathbb X,\!\mathbb X_0,\!\mathbb X_s,\!P,\!\mathbb W,\!F,$ $\!\mathbb Y,\!h)$  with its equivalent transition system  $T(\Sigma)\!=\!(X,\!X_{0},\!X_{S},\!U,\!W,$ $\!\TF,\!Y,\!\mathcal H)$. Suppose $\Sigma_{p}$, $\forall p\!\in\! P$, is $\delta$-ISS as in Definition \ref{def:SFD1}, 
		 with functions $V_p$, $\underline{\alpha}_p, \overline{\alpha}_p, \rho_{p}$ and constant $\kappa_p$,		
		and assume
		Assumptions \ref{ass1} and \ref{ass2} hold. Let $\epsilon>1$. For any design parameters $\varepsilon, \vartheta \in \mathbb R_{\ge 0}$, let $T(\hat{\Sigma})$ be a symbolic model of $T(\Sigma)$ constructed as in Definition \ref{smm}	with any quantization parameter $\eta$ satisfying 
		\begin{align} \label{secquantinit}	
	 \eta \leq \min\{\hat{\gamma}^{-1}((1-\kappa)\varepsilon-\rho(\vartheta)),  \overline{\alpha}^{-1}(\varepsilon)\},
		\end{align}
		where $\kappa \!=\!\max\limits_{p\in P}\left\{\!\kappa^{\frac{\epsilon\!-\!1}{\epsilon}}_{p}\!\right\}$, $\rho\!=\!\max\limits_{p\in P}\left\{{\kappa^{-\frac{k_d}{\epsilon}}_p}\!\!{\rho_{p}}\right\}$, $\hat{\gamma}\!=\!\max\limits_{p\in P}\left\{{\kappa^{-\frac{k_d}{\epsilon}}_p}\!\!{\gamma_{p}}\right\}$, and $\overline{\alpha}\!=\!\max\limits_{p\in P}\left\{{\kappa^{-\frac{l}{\epsilon}}_p}{\overline{\alpha}_p}\right\}$.
		If, $\forall p \in P, ~k_d\geq \epsilon \frac{\ln (\mu)}{\ln (\frac{1}{\kappa_p})}+1$, then function $\mathcal{V}$ defined as 
	 \begin{align}\label{sm}
			\mathcal{V}((x,p,l),(\hat{x},p,l))\Let V_p(x,\hat{x})\kappa^{\frac{-l}{\epsilon}}_p,
			\end{align}
		is a local $\varepsilon$-InitSOPSF from $T(\Sigma)$ to $T(\hat{\Sigma})$.
	\end{theorem}


	Given the results of Theorems \ref{thm:3} and \ref{thm:2}, one can see that conditions \eqref{compoquaninit} and \eqref{secquantinit} may not hold simultaneously. Therefore, we raise the following assumption which provides a small-gain type condition such that one can verify if conditions \eqref{compoquaninit} and \eqref{secquantinit} can be satisfied simultaneously.
	
	\begin{assumption}\label{assump}
		Consider network $\mathcal{I}_{0_N}\!(T(\Sigma_1),\!\ldots,$ $\!T(\Sigma_N))$ induced by $N\!\in\!\N_{\ge1}$ transition systems~$T(\Sigma_i)$. Assume that each $T(\Sigma_i)$ and its symbolic model $T(\hat \Sigma_i)$ admit a local $\varepsilon_i$-InitSOPSF $\mathcal{V}_i$ as in \eqref{sm}, associated with functions and constants $\kappa_i$, $\alpha_i$, and $\rho_{i}$ appeared in Theorem \ref{thm:2}. Define
	 \begin{align}\label{gammad}
			\gamma_{ij}&\Let\left\{
			\begin{array}{lr}
			\!\!(1-\!\kappa_{i})^{-1}\!\rho_i\!\circ\!\alpha_{j}^{-1} &\mbox{if } j \!\in\! \mathcal{N}_i, \\
			\!\!0 &\mbox{otherwise},
			\end{array}\right.
			\end{align}for all $ i,j \in [1;N]$,
		and assume that functions $\gamma_{ij}$ defined in \eqref{gammad} satisfy
		\begin{align}\label{SGC}
			\gamma_{i_1i_2}\circ\gamma_{i_2i_3}\circ\cdots\circ\gamma_{i_{r-1}i_r}\circ\gamma_{i_ri_1}<\mathcal{I}_d,
			\end{align}
		$\forall(i_1,\ldots,i_r)\in\{1,\ldots,N\}^r$, where $r\in \{1,\ldots,N\}$.
	\end{assumption}
	
	The next theorem is main result of the paper. We show that under the above small-gain assumption, one can always compositionally design local quantization parameters to satisfy conditions \eqref{compoquaninit} and \eqref{secquantinit} simultaneously.
	
	\begin{theorem} \label{smallgain}
		Suppose that Assumption \ref{assump} holds. Then, there always exist local quantization parameters $\eta_i$ and $\phi_{ij}$, $\forall i,\!j \!\in\! [1;N]$,  such that \eqref{compoquaninit} and \eqref{secquantinit} can be satisfied simultaneously.
	\end{theorem}

	\section{Case Study}
	Consider a network of discrete-time switched system $\Sigma=(\mathbb X,\mathbb X_0,\mathbb X_s,P,F,\mathbb Y,h)$ as in Definition \ref{netsw}, consisting of $n$ systems $\Sigma_i$ each described by:
	\begin{align}\label{exsm}
	\Sigma_i:\left\{
	\begin{array}[\relax]{rl}
	\mathbf{x}_i(k\!+\!1)\!\!\!\!\!&=a_{i{\mathsf p}_i(k)}\mathbf{x}_i(k)\!+\! d_i\omega_i(k)\!+\!b_{i\mathsf{p}_i(k)},\\
	\mathbf{y}_i(k)\!\!\!\!\!&= c_i \mathbf{x}_i(k),
	\end{array}\right.
	\end{align}
	where $\mathsf p_i (k)\!\in \!P_i \!=\! \{1,2\}$, $k \!\in\! \N$,  denote the modes of each system $\Sigma_i$. The switching signal is set to be $\mathsf{p}_i(k) \!=\! 1$ if $k$ is odd and $\mathsf{p}_i(k) \!=\! 2$ if $k$ is even, $\forall k \!\in\! \N$. The other parameters are as the following: $a_{i1} \!=\! 0.05$, $a_{i2}\! = \!0.1$, $b_{i1}\!=\!0.1$, $b_{i2}\!=\!0.15$, $d_i\!=\!0.05$, $c_i \!=\! [c_{i1};\!\dots\!;c_{in}]$ with $c_{i(i\!+\!1)} \!= \!1$, $c_{ij} \!= \!0$, $\forall i\!\in\! [1;n\!-\!1], \forall j \!\neq\! i\!+\!1$, $c_{n1}\!=\!c_{nn} \!=\! 1$, $c_{nj}\!= \!0$, $\forall j \!\in\! [2;n\!-\!1]$. The internal inputs are subjected to the constraints $\omega_1(k)\! = \!c_{n1}\mathbf{x}_n(k)$ and $\omega_i(k) \!= \!c_{(i\!-\!1)i}\mathbf{x}_{(i\!-\!1)}(k)$, $\forall i\!\in\! [2;n]$.
	For each switched system, the state sets are $\mathbb X_i \!=\!\mathbb X_{0_i} \!=\! (0,\! 0.6)$, $\forall i \!\in\! [1;n]$, the secret sets are $\mathbb{X}_{s_1}\!\! =\!\! (0,\!0.2]$, $\mathbb{X}_{s_2}\!\! =\!\! [0.4,\!0.6)$, $\mathbb{X}_{s_i}\!\! =\!\! (0, \!0.6)$, $\forall i \!\in\! [3;n]$, 
	the output sets are $\mathbb Y_i\!=\!\prod_{j\!=\!1}^n  \!\mathbb Y_{ij}$ where
	$\mathbb Y_{i(i\!+\!1)} \!=\!(0, \!0.6)$, $\mathbb Y_{ii} \!=\!\mathbb Y_{ij} \!=\!\{0\}$, $\forall i \!\in\! [1;n\!-\!1]$, $\forall j \!\neq\! i\!+\!1$, $\mathbb Y_{nn} \!=\!\mathbb Y_{n1} \!=\! (0, \!0.6)$, $\mathbb Y_{nj} \!=\!\{0\}$, $\forall j \!\in\! [2;n\!-\!1]$, and internal input sets are $\mathbb{W}_1\!\!=\!\!\mathbb{Y}_{ni}$, $\mathbb{W}_i\!\!=\!\!\mathbb{Y}_{(i\!-\!1)i}$, $\forall i\!\in\! [2;n]$.
	Intuitively, the output of the network is the external output of the last system $\Sigma_n$. The main goal of this example is to check approximate initial-state opacity of the concrete network using its symbolic model. Now, let us construct a symbolic model of $\Sigma$ compositionally with accuracy $\hat \varepsilon \!=\! 0.25$ as defined in \eqref{er}. We use our compositional approach to achieve this goal.
	
	Consider functions $V_{ip_i}\! \!=\! \!|x_i- \hat x_i|$, $\forall i \!\in\! [1;n]$. It can be readily verified that \eqref{e:SFC11} and \eqref{e:SFC22} are satisfied with 
	$\underline{\alpha}_{ip_i}\!=\!\overline{\alpha}_{ip_i}\!=\!\mathcal{I}_d$, $\rho_{ip_i}\!=\!0.05$, $\forall p_i\!\in\! P_i$,  $\kappa_{i1}\!=\! a_{i1}\!=\!0.05$,  $\kappa_{i2}\!=\! a_{i2}\!=\!0.1$.
	Condition \eqref{tinq} is satisfied with $\gamma_{ip_i}\!=\!\mathcal I_d$, $\forall p_i\!\in\! P_i$. Moreover, since $V_{ip_i}\!= \!V_{iq_i},\forall p_i,q_i\!\in\! P_i$, $V_i(x_i,\hat{x}_i)\!=\! |x_i\!-\!\hat x_i|$ is a common $\delta$-ISS Lyapunov function for system $\Sigma_i$. 
	Next, given functions $\kappa_i \!=\! 0.1$, $\rho_i\!=\!0.06\mathcal I_d$, $\alpha_i\!=\! \mathcal{I}_d$, $\hat{\gamma}_i\!=\!1.05\mathcal I_d$, and $\overline{\alpha}_i\!=\!\mathcal{I}_d$ as appeared in Theorem \ref{thm:2}, we have $\gamma_{ij} \!<\! \mathcal{I}_d$ by \eqref{gammad}, $\forall i,j \!\in\! [1;n]$.
	Hence, the small-gain condition \eqref{SGC} is satisfied. Then, by applying Theorem \ref{smallgain}, we obtain proper pairs of local parameters $(\varepsilon_i,\!\vartheta_i)\!=\! (0.25, \!0.25)$ for all of the transition systems. 
	Accordingly, we provide a suitable choice of local quantization parameters as $\eta_i=\!0.2$, $\forall i \!\in\! [1;n]$, such that inequality \eqref{secquantinit} for each transition system $T( {\Sigma}_i)$ is satisfied. 
	Then, we construct local symbolic models $T(\hat{\Sigma}_i)\!=\!(\hat{X}_i,\!\hat X_{0_i},\!\hat X_{s_i},\!\hat{U}_i,\!\hat{W}_i,\!\hat{\TF}_i,\!\hat{Y}_i,\!\hat{\op}_i)$ as defined in Definition \ref{smm},  where $\hat{X}_i\!=\! \hat{{X}}_{0_i} \!=\! \{0.2,\!0.4\}\! \times\! \{1,2\} \!\times\! \{0\}$, $\hat{X}_{s_1}\!=\! \{0.2\}\! \times\! \{1,2\} \!\times\! \{0\}$, $\hat{X}_{s_2}\!=\! \{0.4\}\! \times\! \{1,2\} \!\times\! \{0\}$, $\hat{X}_{s_i}\!=\! \{0.2,\!0.4\}\! \times\! \{1,2\}\!\times\! \{0\}$, $\forall i \!\in\! [3;n]$, $\hat Y_{i}=\!\!\prod_{j\!=\!1}^{i} \!\{0\} \!\times\!\!\{0.2,\!0.4\}\!\times\!\!\prod_{j\!=\!i+2}^{n}\{0\}$, $\forall i \!\in\! [1;n\!-\!1]$, $\hat Y_{n}\!=\!\{0.2,\!0.4\} \!\times\! \!\prod_{j\!=\!2}^{n\!-\!1}\!\{0\} \times\!\{0.2,\!0.4\}$, $\hat W_i\!=\!\{0.2,\!0.4\} \!$, $\forall i \!\in\! [1;n]$.  
	Now, using the result in Theorem \ref{thm:2}, one can verify that $V_i((x_i,\!p_i,\!l_i),\!(\hat{x}_i,\!p_i,\!l_i))\!=\! |x_i\!-\!\hat x_i|$ is a local $\varepsilon_i$-InitSOPSF from each $T(\Sigma_i)$ to its symbolic model $T(\hat{\Sigma}_i)$. Furthermore, by the compositionality result in Theorem \ref{thm:3}, we obtain that ${V} \!\!=\!\! \max\limits_{i}\{V_i((x_i,\!p_i,\!l_i),\!(\hat{x}_i,\!p_i,\!l_i)) \} \!\!=\!\! \max\limits_{i}\{|x_i\!-\! \hat x_i|\}$ is an $\varepsilon$-InitSOPSF from $T(\Sigma)\!\!=\!\!\mathcal{I}_{0_N}(T(\Sigma_1),\!\ldots\!,T(\Sigma_{N}))$ to $T(\hat{\Sigma})\!\!=\!\!{\mathcal{I}}_{0_N}(T(\hat{\Sigma}_1),\!\ldots\!,T(\hat{\Sigma}_{N}))$ with  $\varepsilon \!=\! \max\limits_{i} \varepsilon_i\! =\! 0.25$.
		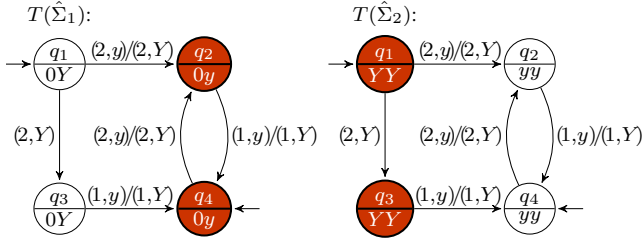
\begin{figure}[t!]
			\vspace{0.2cm}
		\centering
		{
			\resizebox{9cm}{!}{
				\footnotesize{
					\begin{tikzpicture}[->,>=stealth',shorten >=1pt,auto,node distance=2cm, inner sep=1pt, initial text =,
					every state/.style={draw=black,fill=white,state/.style=state with output},
					accepting/.style={draw=black,thick,fill=red!80!green,text=white},bend angle=25]

					\node[] at (-1,0.7) {$T(\hat \Sigma_1)$:};
					
					\node[] at (3.5,0.7) {$T(\hat\Sigma_2)$:};

					\node[state with output, initial] (A1)         at (-1,0)              {$q_1$ \nodepart{lower} $0Y$};
					\node[state with output, accepting]         (B1) [right of=A1] {$q_2$ \nodepart{lower} $0y$};
					\node[state with output]         (C1) [below of=A1] {$q_3$ \nodepart{lower} $0Y$};
					\node[state with output, accepting,  initial right]         (D1) [below of=B1]       {$q_4$ \nodepart{lower} $0y$};

					\path (A1) edge    node {$(\!2,\!y\!)\!/\!(\!2,\!Y\!)\!$} (B1)
					edge              node  [left] {$(\!2,\!Y\!)$} (C1)
					(B1) edge      [bend left]          node {$(\!1,\!y)\!/\!(\!1,\!Y\!)$} (D1)
					(C1) edge             node {$(\!1,\!y)\!/\!(\!1,\!Y\!)$} (D1)
					
					(D1) 	edge     [bend left]       node  {$(\!2,\!y\!)\!/\!(\!2,\!Y\!)$} (B1)

					;

					\node[state with output, accepting,  initial] (A2)       at (3.5,-0)             {$q_1$ \nodepart{lower} $YY$};
					\node[state with output]         (B2) [right of=A2] {$q_2$ \nodepart{lower} $yy$};
					\node[state with output, accepting]         (C2) [below of=A2] {$q_3$ \nodepart{lower} $YY$};
					\node[state with output, initial right]         (D2) [below of=B2]       {$q_4$ \nodepart{lower} $yy$};

					\path (A2) edge    node {$(\!2,\!y\!)\!/\!(\!2,\!Y\!)\!$} (B2)
					edge             node  [left] {$(\!2,\!Y\!)$} (C2)
					(B2) edge      [bend left]  node {$(\!1,\!y)\!/\!(\!1,\!Y\!)$} (D2)
					(C2) edge             node {$(\!1,\!y)\!/\!(\!1,\!Y\!)$} (D2)
					
					(D2) 	edge     [bend left]   node  {$(\!2,\!y\!)\!/\!(\!2,\!Y\!)$} (B2)
					;	
					\end{tikzpicture}
				}
		}}
		\caption{Local symbolic models of transition systems.}
		\label{exautomata1}
		\vspace{-0.5cm}
	\end{figure}
	
	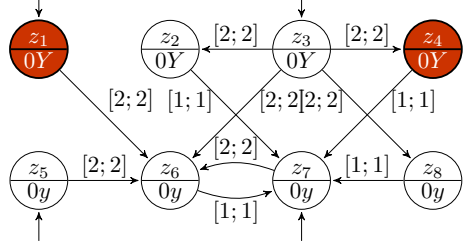
\begin{figure}[t!]
		\centering
		{
			\resizebox{7.0cm}{!}{
				\begin{tikzpicture}[->,>=stealth',shorten >=1pt,auto,node distance=2.0cm, inner sep=1pt, initial text =,
				every state/.style={draw=black,fill=white,state/.style=state with output},
				accepting/.style={draw=black,thick,fill=red!80!green,text=white},bend angle=20]

				\node[] at (0.5,-0) {$\mathcal{I}(T(\hat \Sigma_1),\!T(\hat \Sigma_2))$:};
				\node[] at (2.8,-1.9) {$[1;1]$};
				\node[] at (4.2,-1.9) {$[2;2]$};
				\node[] at (4.8,-1.9) {$[2;2]$};
				
				\node[] at (6.2,-1.9) {$[1;1]$};

				\node[state with output,initial above, accepting]         (A) at (0.5,-1.1)  {$z_1$ \nodepart{lower} $0Y$};
				\node[state with output] (B)       [right of=A]          {$z_2$ \nodepart{lower} $0Y$};
				\node[state with output,initial above]  (C) [right of=B]       {$z_3$ \nodepart{lower} $0Y$};
				\node[state with output, accepting]         (D) [right of=C] {$z_4$ \nodepart{lower} $0Y$};
				\node[state with output,initial below]         (E) [below of=A] {$z_5$ \nodepart{lower} $0y$};	
				\node[state with output]         (F) [right of=E]       {$z_6$ \nodepart{lower} $0y$};
				\node[state with output,initial below]         (G) [right of=F] {$z_7$ \nodepart{lower} $0y$};
				\node[state with output]         (H) [right of=G]       {$z_8$ \nodepart{lower} $0y$};

				\path
				(A) edge      node {$[2;2]$} (F)
				(B) edge         node  {} (G)
				(C) edge     node  {} (F)
				(C) edge      node [above]{$[2;2]$} (B)
				(C) edge      node {} (H)
				(C) edge      node {$[2;2]$} (D)
				(D) edge      node {} (G)
				(E) edge          node {$[2;2]$} (F)
				(F) edge      [bend right]        node[below] {$[1;1]$} (G)
				
				(G) edge    [bend right]          node[above] {$[2;2]$} (F)
				
				(H) edge    node[above] {$[1;1]$} (G);

				\end{tikzpicture}
			}
		}
		\caption{Symbolic model of a network of 2 transition systems.}
		\label{exautomata2}
		\vspace{-0.5cm}
	\end{figure}

	Now, let us verify approximate initial-state opacity for $T(\Sigma)$ using the network of symbolic models $T(\hat{\Sigma})$. To do this, we first show an example of a network consisting of $2$ transition systems, as shown in Figures~\ref{exautomata1} and \ref{exautomata2}. The two automata in Figure~\ref{exautomata1}  represent the symbolic models of the local transition systems, and the one in Figure~\ref{exautomata2} is the network of symbolic models. Each circle is labeled by the state (top half) and the corresponding output (bottom half). Initial states are distinguished by being the target of a sourceless arrow. The symbols on the edges show the switching signals $\!\mathsf p_i (k) \!\in\! \{1,2\}$ and internal inputs coming from other local transition systems. For simplicity of demonstration, we use symbols to represent the state and output vectors, where $q_1\!=\![0.4,\!2,\!0]$, $q_2\!=\![0.2,\!1,\!0]$,  $q_3\!=\![0.4,\!1,\!0]$,  $q_4\!=\![0.2,\!2,\!0]$,  $z_1\!=\![q_4;\!q_1]$, $z_2\!=\![q_3;\!q_3]$, $z_3\!=\![q_1;\!q_1]$, $z_4\!=\![q_2;\!q_3]$, $z_5\!=\![q_1;\!q_4]$, $z_6\!=\![q_2;\!q_2]$, $z_7\!=\![q_4;\!q_4]$, $z_8\!=\![q_3;\!q_2]$, $y\!=\!0.2$, $Y\!=\!0.4$, $0y \!=\! [0;\!0.2]$, $0Y \!=\! [0;\!0.4]$, $yy \!=\! [0.2;\!0.2]$, $YY \!=\! [0.4;\!0.4]$.  
	One can easily see that $\mathcal{I}_{0_N}(T(\hat{\Sigma}_1),\!T(\hat{\Sigma}_2))$ is $0$-approximate initial-state opaque, since for any run starting from any secret state, i.e. $z_1$ and $z_4$, there exists a run from a non-secret state, i.e. $z_2$ and $z_3$, such that the output trajectories are exactly the same. One can readily verify that the symbolic network has this property regardless of the number of systems (i.e. $n$), due to the homogeneity of systems $\Sigma_i$ and the symmetry of the circular network topology. Thus, one can conclude that $T(\hat \Sigma) \!\!=\!\! \mathcal{I}_{0_N}(T(\hat{\Sigma}_1),\!\dots\!,T(\hat{\Sigma}_n))$ is $0$-approximate initial-state opaque. Therefore, by Corollary \ref{thm:InitSOP}, we obtain that the original network $T(\Sigma) \!\!=\! \!\mathcal{I}_{0_N}(T(\Sigma_1),\!\dots\!,T(\Sigma_n))$ is $0.5$-approximate initial-state opaque.

	\bibliographystyle{ieeetr}
	\bibliography{refr6}      
\end{document}